# High-performance coherent population trapping atomic clock with direct-modulation distributed Bragg reflector laser

Peter Yun[1], Qinglin Li[1,2], Qiang Hao[1], Guobin Liu[1], Emeric de Clercq[3], Stéphane Guérandel[3], Xiaochi Liu[4], Sihong Gu[5], Yuping Gao[1], and Shougang Zhang[1]

[1] National Time Service Center, Chinese Academy of Sciences, Xi'an, China
[2] School of Astronomy and Space Science, University of Chinese Academy of Sciences, Beijing, China
[3] LNE-SYRTE, Observatoire de Paris, Paris, France
[4] National Institute of Metrology, Beijing, China
[5] Innovation Academy for Precision Measurement Science and Technology, Chinese Academy of Sciences, Wuhan, China
E-mail: yunenxue@ntsc.ac.cn

**Abstract**

The coherent population trapping (CPT) atomic clock is very promising for use in next-generation spaceborne applications owing to its compactness and high performance. In this paper, we propose and implement a CPT atomic clock based on the direct modulation of a large-modulation-bandwidth and narrow-linewidth distributed Bragg reflector laser, which replaces the usually used external bulk modulator in the high-performance CPT clock. Our method retains the high performance while significantly reducing the size.

Using this highly compact bichromatic light source and simplest CPT configuration, in which a circularly polarized bichromatic laser interrogates the $^{87}$Rb atom system, a CPT signal of clock transition with a narrow linewidth and high contrast is observed. We then lock the local oscillator frequency to the CPT error signal and demonstrate a short-term frequency stability of $3.6 \times 10^{-13} \tau^{-1/2}$ (4 s $\leqslant \tau \leqslant$ 200 s). We attribute it to the ultralow laser frequency and intensity noise as well as to the high-quality-factor CPT signal. This study can pave the way for the development of compact high-performance CPT clocks based on our scheme.

(Some figures may appear in colour only in the online journal)

## 1. Introduction

The recent progress on high-performance coherent population trapping (CPT) [1, 2] atomic clocks with excellent short-term [3–6] and mid-term [7, 8] frequency stabilities paves the way for deep space exploration, next-generation Global Navigation Satellite System, high-speed communication, etc. However, they need to be more compact and robust for broader applications.

In CPT atomic clocks, a coherent bichromatic laser is needed to couple two ground states to a common excited state and prepare the atom system to the CPT states. Simple and sophisticated methods have been proposed to generate a bichromatic laser. Both have advantages and disadvantages. The simplest approach to generate a coherent bichromatic laser is to directly modulate a vertical-cavity surface-emitting laser [9, 10]. However, besides its large relative intensity noise (RIN), its relatively broad linewidth (~100 MHz) would also induce a large frequency modulation (FM) to amplitude modulation (AM) noise through the atom resonance line. Thus, high-performance CPT atomic clocks usually utilize more complex methods such as optical phase locking of two independent lasers in lin⊥lin CPT [11–13] or inject locking [14, 15]. However, the two-laser system in both schemes hinders compact and robust applications. External modulators, such as acousto-optic modulator (AOM) or electro-optic modulator, are usually employed in push-pull CPT [16, 17], lin//lin CPT [18–20], double-modulation (DM) CPT [21, 22], and lin⊥lin CPT [23]. Excellent short-term frequency stabilities at the level of $1–3 \times 10^{-13} \tau^{-1/2}$ have been typically observed [3–6]. However, these methods increase the size, weight, power, cost, and complexity of the CPT clock.

In this study, we demonstrate a clock in which we directly modulate a distributed Bragg reflector (DBR) laser diode and apply it to the simplest CPT configuration (denoted as (σ, σ) CPT). An encouraging short-term frequency stability is observed. Thus, it is feasible to build a highly compact atomic clock. This method could also be applied to cold-atom CPT clocks [24–27].

## 2. Experimental setup and preparation

### 2.1 Experimental setup

Our setup is depicted in figure 1, which is composed roughly of three parts, the DBR laser, which generates a multi-chromatic light, reference cell for laser frequency locking, and clock $^{87}$Rb cell for local oscillator (LO) locking.

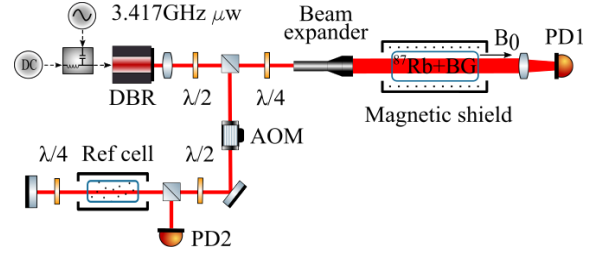

**Figure 1.** Optical setup.

A DBR laser (Photodigm, PH795DBR) with a wavelength of 795 nm, linewidth of approximately 1 MHz, and output optical power up to 80 mW is used as a light source. Its temperature is regulated at 24 °C. Then we drive it with a current source (Vescent, D2-105) and tune its wavelength to the working point of the CPT resonance for the $D_1$ line of $^{87}$Rb with a current of approximately 100 mA. To generate a bichromatic light for the CPT experiment, a 3.417-GHz microwave signal is coupled to the laser diode current through a bias-tee. As demonstrated in our previous study [28], with a typical microwave power of 22–28 dBm, we can obtain the maximum efficiency (~50%) of laser power transfer to the ±1st sidebands, which form the desired bichromatic light for CPT interaction.

We lock our laser frequency with a simple scheme [29], in which an AOM is used to compensate the buffer-gas-induced optical frequency shift in the CPT clock cell.

We then use a buffer-gas vapor cell and simplest CPT scheme with a circularly polarized light beam to demonstrate a high-performance CPT clock. The laser beam is expanded to a circular beam with a diameter of approximately 18 mm before passing through the vapor cell. The $^{87}$Rb isotope enriched cylindrical vapor cell (diameter: 20 mm, length: 50 mm) is filled by a buffer gas mixture of argon and nitrogen with a total pressure of 25 Torr. The cell temperature is stabilized to approximately 62 °C. Unless otherwise specified, a uniform magnetic field of $B_0 = 3.66$ μT is applied along the direction of the cell axis by means of a solenoid to remove the Zeeman degeneracy. The ensemble is surrounded by two magnetic shields to reject Earth and other stray magnetic fields.

### 2.2 Microwave phase noise

Before we describe the optical experiment, we briefly introduce our microwave source, which is crucial for the CPT clock. With the microwave chain, shown in the insert of figure 2, similar to that in ref [30], we built a 3.417-GHz microwave source with an ultralow phase noise. The absolute phase noise of the 3.417-GHz signal is measured to be -122.5 dBc/Hz at an offset frequency of 286 Hz. Compared to the ideal frequency multiplication, it exhibits a deterioration of only 1 dB at the second harmonic of our atomic clock modulation frequency. The phase noise contributions to the CPT clock frequency stability are estimated in table I.

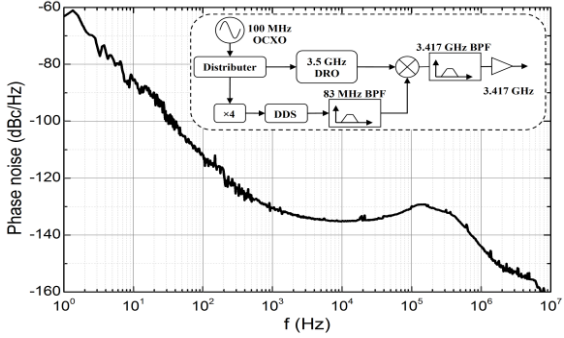

**Figure 2.** Phase noise of the 3.417-GHz microwave signal; the inset shows the microwave chain.

*2.3 Laser frequency and intensity noise*

We use the sub-Doppler absorption to lock the laser frequency with the dual-frequency laser field method [29]. In this setup, we split a part of the multi-chromatic laser with an optical power of 1.8 mW and beam diameter of 3 mm to the reference cell, which is temperature-stabilized at 24 °C. The measured spectra in figure 3 correspond to the $D_1$ line of the $^{87}$Rb spectrum in the vacuum reference cell and clock cell recorded with the multi-chromatic laser. The optical transitions in the clock cell are broadened and shifted by collisions between the Rb atoms and buffer gas molecules. The frequency shift (~222 MHz) is already compensated by the AOM in this plot.

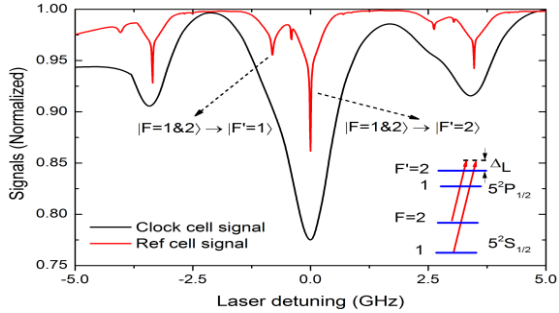

**Figure 3.** Spectra of the $^{87}$Rb $D_1$ line in the vacuum reference cell and clock cell recorded with the multi-chromatic laser. The distortion with a laser detuning lower than approximately -3.7 GHz is caused by the limited mode-hopping-free tuning ranges of the diode laser. (Inset) Atomic levels involved in the $D_1$ line of rubidium.

With the narrow absorption peak and high signal-to-noise ratio, as shown in figure 3, it is simple to lock the laser frequency, without unlocking for days. The laser frequency noise suppression with this method compared to the free-running laser (figure 4a) is impressive. Its contribution to the CPT clock's short-term frequency stability is negligible through the FM–AM effect, see table I.

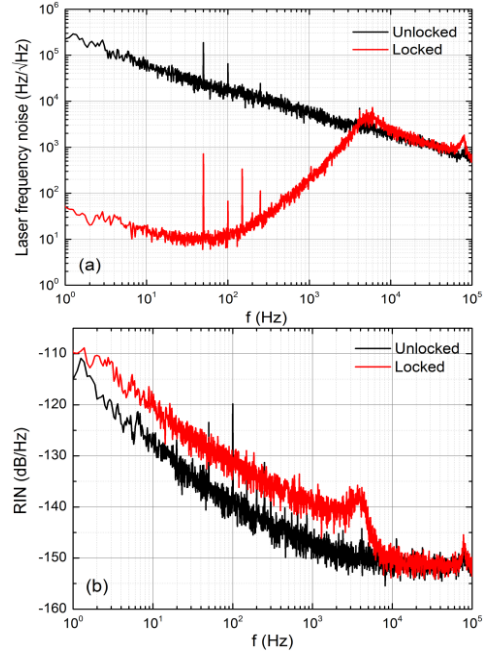

**Figure 4.** Laser frequency (a) and intensity (b) noises with and without laser frequency locking.

The laser relative intensity noise (RIN) is also measured and plotted in figure 4b. The laser RIN is deteriorated by the laser frequency locking. The peak frequency is approximately 4 kHz to 5 kHz, equal to the feedback bandwidth of the laser frequency locking, as shown in figure 4a. It is of interest to investigate the physics behind this notable phenomenon in future studies.

## 3. Experimental results

*3.1 Spectroscopy studies*

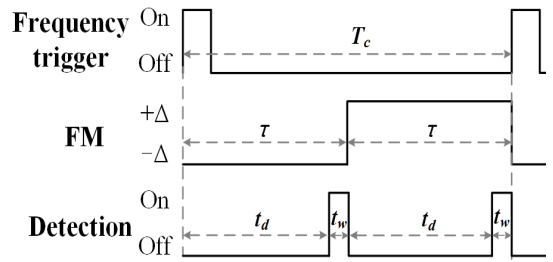

**Figure 5**. Time sequence with a total time cycle of $T_c$, FM half-period of $\tau = T_c/2$, detection window of $t_w$, and delay time of $t_d = \tau - t_w$.

For the simplest CPT scheme, analog or digital modulation/demodulation methods could be applied to obtain the CPT signal and error signal, and then lock the LO. To optimize the CPT contrast (defined as the CPT amplitude to background ratio) and linewidth, we use the digital method with the time sequence shown in figure 5 to perform the CPT experiment.

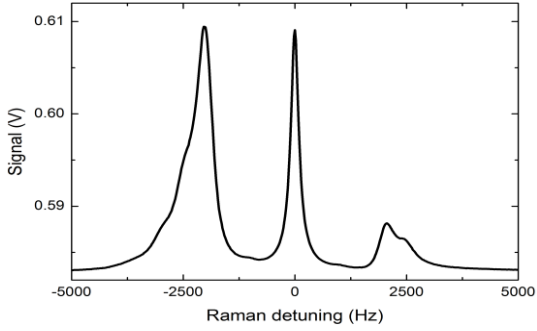

**Figure 6.** Zeeman spectrum. The center peak is the 0–0 clock transition. The working parameters are $T_c = 7$ ms, $t_w = 1$ ms, laser power $P_L = 0.654$ mW, microwave power $P_{\mu w} = 26.9$ dBm, and $B_0 = 146.4$ nT.

We observed the CPT signals with this time sequence, showing all allowed CPT transitions between Zeeman sublevels of the $^{87}$Rb ground state (figure 6). The amplitude of clock transition is not the most pronounced as in the ($\sigma^-$, $\sigma^-$) CPT scheme due to optical pumping towards the lowest Zeeman sublevels. The broadening and distortion of neighboring lines are caused by magnetic-field inhomogeneities.

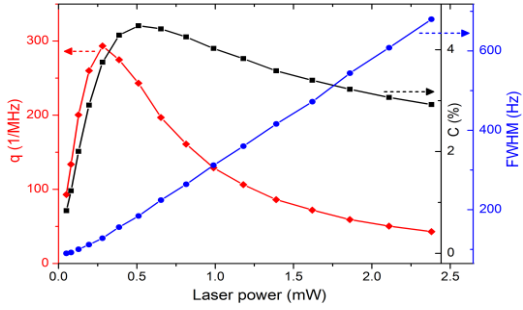

**Figure 7.** Contrast ($C$), linewidth (full width at half maximum (FWHM)), and quality factor $q = C$/FWHM versus the laser power. The working parameters are $T_c = 7$ ms, $t_w = 1$ ms, and $P_{\mu w} = 26.9$ dBm.

We then studied the CPT clock transition's contrast ($C$), linewidth (FWHM), and quality factor $q = C$/FWHM as functions of the laser power ($P_L$) and microwave power ($P_{\mu w}$), plotted in figures 7 and 8, respectively. In these measurements, $T_c$ and $t_w$ are already optimized for maximizing contrast. With the guide of these optimal parameters, we obtained a CPT signal and error signal (figure 9) with FWHM = 156 Hz and contrast of 4.3%, comparable to the values obtained in ref [5], which provides a good base for clock applications.

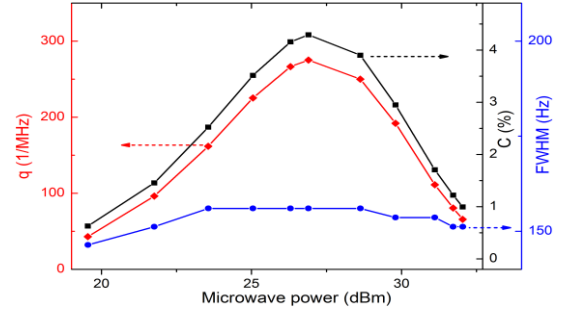

**Figure 8.** Contrast, FWHM, and $q$ versus the 3.417-GHz microwave power. The working parameters are $T_c = 7$ ms, $t_w = 1$ ms, and $P_L = 0.387$ mW.

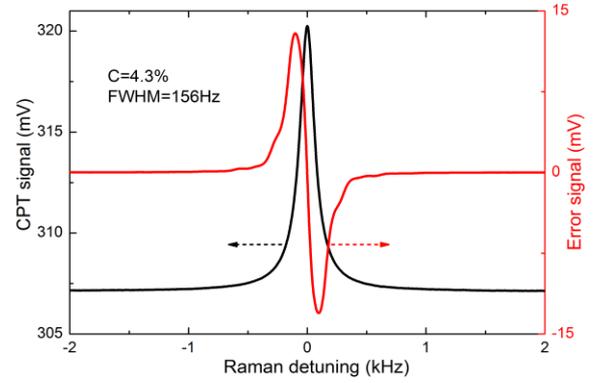

**Figure 9.** Typical CPT and error signal of the clock transition. The working parameters are $T_c = 7$ ms, $t_w = 1$ ms, $P_L = 0.387$ mW, and $P_{\mu w} = 26.9$ dBm.

*3.2 Short-term frequency stability and noise budget*

With the observed CPT error signal, we can lock our LO to the CPT clock transition frequency. The Allan standard deviations of the unlocked and locked LO frequencies, measured against our H maser, are shown in figure 10.

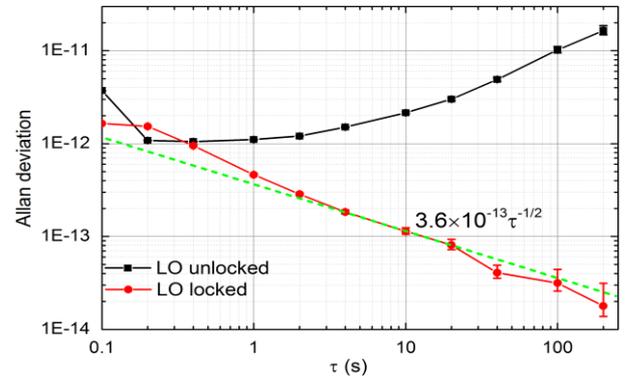

**Figure 10.** Allan deviation of the LO frequency stability: free running (black) and locked (red) on the atomic resonance; the slope of the green dashed guide line is $3.6 \times 10^{-13} \tau^{-1/2}$.

The locked LO frequency stability is $3.6 \times 10^{-13} \tau^{-1/2}$ with averaging time from 4 s to 200 s, and it reaches $1.8 \times 10^{-14}$ at 200 s. This encouraging stability paves the way for a compact high-performance CPT clock.

The laser intensity noise after the interaction with the atomic vapor is shown in figure 11. It shows the different contributions to the amplitude noise, i.e., the detector noise, shot noise, laser FM–AM, and laser AM–AM noise. The noise spectral density is 140.8 nV/√Hz at the Fourier frequency of 143 Hz, i.e., the microwave modulation frequency, which leads to an Allan deviation of $1.02 \times 10^{-13}$ at 1 s. According to table I, this value is very close to the quadratic sum of the individual contributions.

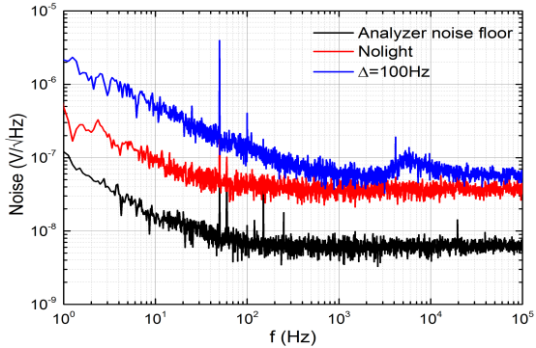

**Figure 11.** RIN after the cell, detector noise, and analyzer noise floor. The working parameters are the same as in figure 9.

With the method in ref [5], we estimate the contributions of the various noise sources to the short-term instabilities, as summarized in Table I. The quadratic sum of all noise contributions leads to an Allan deviation at 1 s of $1.08 \times 10^{-13}$. The dominant contribution is the laser RIN, which could be reduced by active laser power locking or by overcoming the RIN degradation by the laser frequency locking mentioned above.

**TABLE I.** Noise contributions to the stability at 1 s.

| Noise term | Noise level | $\sigma_y$ (1 s) $\times 10^{13}$ |
|---|---|---|
| **Detector noise** | 36.8 nV/√Hz | 0.266 |
| **Shot noise** | 38.8 nV/√Hz | 0.281 |
| **FM–AM** | 1.4 nV/√Hz | 0.010 |
| **AM–AM** | 125.3 nV/√Hz | 0.905 |
| **AM–FM** | 1.3 nW@1s | 0.070 |
| **FM–FM** | ~50 Hz@1s | <0.001 |
| **μw phase noise** | -122.5 dBc/Hz | 0.442 |
| **Total** | | 1.081 |

The measured short-term stability is larger than the predicted value. The disagreement is believed to be induced by the microwave-power fluctuations [5]. This term could be reduced by microwave power stabilization or well-chosen laser power, referred to as "magic power point" [31, 32].

As we focused on the high-quality-factor CPT signal and potential for a high-performance CPT clock, the mid-term and long-term frequency stabilities were not fully investigated. We aim to evaluate them after integration of the table experiment setup into a compact clock.

## 4. Outlook and summary

To the best of our knowledge, this is the first report on a CPT clock based on a direct-modulation diode laser to reach a short-term frequency stability of $3.6 \times 10^{-13} \tau^{-1/2}$ (4 s ≤ τ ≤ 200 s), which is attributed mainly to the ultralow laser frequency and intensity noise as well as the narrow linewidth and high-quality-factor CPT signal.

This progress may provide three opportunities for CPT clocks. First, our scheme could be applied to a CPT configuration with even higher contrast and quality factor, such as push-pull CPT, lin//lin CPT, DM CPT, and Ramsey-CPT, which could improve the CPT clock frequency stability to an even higher level. Second, the microwave power consumption in our scheme is comparable to that of the external modulator method [5]. However, with the DBR diode technology development, optimization of the impedance match, and cavity-enhanced modulation efficiency method [33, 34], we could be able to remarkably reduce the microwave power. This point is crucial to implement a compact high-performance CPT atomic clock. Third, the bulk AOM used in our laser locking setup could be omitted with a paraffin-coated clock cell [35] or even the laser locking setup can be abandoned by using the clock cell. We can lock the laser frequency as usually implemented in a chip-scale atomic clock (CSAC). Thus, the laser can be adopted in CSAC and boost its performance.

### Acknowledgements

We thank Moustafa Abdel Hafiz and Rodolphe Boudot from FEMTO-ST for helpful discussions and manuscript polish. This work was funded by the National Natural Science Foundation of China (grant no. U1731132).